# 3D Fusion between Fluoroscopy Angiograms and SPECT Myocardial Perfusion Images to Guide Percutaneous Coronary Intervention

Haipeng Tang, MS[1#], Robert R. Bober, MD[2#], Chen Zhao, MS[3], Chaoyang Zhang, PhD[1], Huiqing Zhu, PhD[1], Zhuo He, BS[3], Zhihui Xu, MD[4]*, Weihua Zhou, PhD[3]*

1. School of Computing Sciences and Computer Engineering, University of Southern Mississippi, Hattiesburg, MS, 39406, USA
2. Department of Cardiology, Ochsner Medical Center, New Orleans, LA, 70121, USA
3. College of Computing, Michigan Technological University, Houghton, MI, 49931, USA
4. Department of Cardiology, The First Affiliated Hospital of Nanjing Medical University, Nanjing, 210000, China

Short title: Image Fusion to Improve PCI

Haipeng Tang and Robert Bober contributed equally to this work.

Authors' emails:

| | |
|---|---|
| Haipeng Tang, MS | E-mail: haipeng.tang@usm.edu |
| Robert R. Bober, MD | E-mail: rbober@ochsner.org |
| Chen Zhao, MS | E-mail: chezhao@mtu.edu |
| Chaoyang Zhang, PhD | E-mail: chaoyang.zhang@usm.edu |
| Huiqing Zhu, PhD | E-mail: huiqing.zhu@usm.edu |
| Zhuo He, BS | E-mail: zhuoh@mtu.edu |

*Corresponding Authors:

Weihua Zhou, PhD     E-Mail: whzhou@mtu.edu
College of Computing, Michigan Technological University, Houghton, MI, 49931, USA
Tel: 906-487-2666

Or

Zhihui Xu, MD     E-mail: wx_xzh@njmu.edu.cn
Department of Cardiology, The First Affiliated Hospital of Nanjing Medical University, Nanjing, 210000, China;
Tel: (+86)02568303120   or   (+86)15996273639

# Abstract

**Background.** Percutaneous coronary intervention (PCI) in stable coronary artery disease (CAD) is commonly triggered by abnormal myocardial perfusion imaging (MPI). However, due to the possibilities of multivessel disease, serial stenoses and variability of coronary artery perfusion distribution, opportunity exists to better align anatomic stenosis with perfusion abnormalities to improve revascularization decisions. This study aims to develop a 3-dimensional (3D) multi-modality fusion approach to assist decision-making for PCI.

**Methods.** Coronary arteries from fluoroscopic angiography (FA) were reconstructed into 3D artery anatomy. Left ventricular (LV) epicardial surface was extracted from single-photon emission computed tomography (SPECT). The 3D fusion between artery anatomy and LV epicardial surface was completed with scaling iterative closest points (S-ICP) and vessel-surface overlay algorithms. The accuracy of the 3D fusion was evaluated via both computer simulation and real patient data. For technical validation, simulated FA and MPI were integrated and then compared with the ground truth from a digital phantom. For clinical validation, FA and SPECT images were integrated and then compared with the ground truth from computed tomography (CT) angiograms.

**Results.** In the technical evaluation, the distance-based mismatch error between simulated fluoroscopy and phantom arteries is 1.86±1.43mm for left coronary arteries (LCA) and 2.21±2.50mm for right coronary arteries (RCA). In the clinical validation, the distance-based mismatch errors between the fluoroscopy and CT arteries were 3.84±3.15mm for LCA and 5.55±3.64mm for RCA. The presence of the corresponding fluoroscopy and CT arteries in the AHA 17-segment model agreed well with a Kappa value of 0.91(95% confidence interval (CI): 0.89-0.93) for LCA and a Kappa value of 0.80 (CI: 0.67-0.92) for RCA.

**Conclusions.** Our fusion approach is technically accurate to assist PCI decision-making and is clinically feasible to be used in the catheterization laboratory. There is an opportunity to improve the decision-making and outcomes of PCI in stable CAD.

**Abbreviations:**

CAD = Coronary artery disease
DP = Dynamic programming
DP-LV = Left ventricular epicardial surface extracted by dynamic programming-based method
FA = Fluoroscopy angiography
LV = Left ventricle
LCA = Left coronary arteries
LAD = Left anterior descending
LCX = Left circumflex artery
ME-LV = Left ventricular epicardial surface extracted by manually drawing
MPI = Myocardial perfusion imaging
PCI = Percutaneous coronary intervention
PDA = Posterior descending artery
PLB = Posterolateral branch artery
RCA = Right coronary arteries
SPECT = Single-photon emission computed tomography
S-ICP = Scaling iterative closest points
X-CAT = Extended cardiac-torso phantom

## 1. INTRODUCTION

In stable coronary artery disease (CAD), mortality and morbidity benefits of revascularization by percutaneous coronary intervention (PCI) have not been fully realized in clinical trials[1–4]. Several hypotheses exist to explain the findings of these clinic trials. One hypothesis is revascularization, although visually "successful", does not improve myocardial perfusion because incorrect lesions and/or vessel(s) are targeted. This is very plausible especially in cases with multivessel disease or serial stenoses. Typically, fluoroscopic angiography (FA) is performed independently of functional data such as myocardial perfusion imaging (MPI) and therefore, the image datasets are clinically segregated. In addition, the lack of patient-specific anatomy on functional datasets reduces its target specificity especially in cases with serial stenosis, multivessel disease or coronary anomalies. Hypothetically, individualized registration of FA and MPI datasets could assist and improve revascularization decisions if anatomic and functional abnormalities could be accurately aligned.

To test this hypothesis several processes must be developed. First, 2D FA datasets must be accurately converted into 3D datasets while maintaining anatomic precision. Second, extraction of left ventricular (LV) epicardial surface from MPI datasets must be accurate. Third, and most importantly, fusion of the 3D FA datasets with LV MPI datasets must be accurate. Fourth, the conversion, extraction, and fusion processes must be fast enough such workflow is not compromised and revascularization is not delayed.

Several fusion techniques, landmark-based[5,6] and rigid iterative closest points (ICP)[7,8], were developed and validated over the past decade. They require a pair of size-matched 3D artery anatomy from fluoroscopy angiograms and LV surface from single-photon emission computed tomography (SPECT) images. This condition is quite difficult to meet because of heart beating and thus image acquisitions at different cardiac frames, which affects the accuracy of the 3D fusion between artery anatomy and LV surface. In order to match the time points, several studies used principal component analysis (PCA) based[8] or visual estimation-based methods[6] to select and fuse the end-diastolic fluoroscopy angiograms and end-diastolic SPECT images. However, all these estimation-based methods cannot guarantee a pair of size-matched artery anatomy and LV surface. A deformable registration algorithm is needed to improve the accuracy of the 3D fusion.

The objective of this study was to develop a deformable 3D fusion approach to integrate 3D coronary artery anatomy from fluoroscopy angiograms with LV epicardial surface from SPECT MPI to guide revascularization decision-making.

## 2. METHODS

First, 3D arterial anatomy was reconstructed from fluoroscopy angiograms via imaging geometry calibration and vessel reconstruction algorithms. Second, LV epicardial surface was extracted from SPECT MPI images using a dynamic programming-based algorithm. Third, the 3D artery anatomy was registered with the LV epicardial surface using scaling iterative closest points (S-ICP) algorithm and then overlaid onto the surface using a vessel-surface overlay algorithm. A computer simulation was executed to technically evaluate the accuracy of the 3D fusion approach. Real patient data was used to evaluate the clinical feasibility of the 3D fusion approach.

### 2.1 Fluoroscopy Image Processing

### 2.1.1. Reconstruction of 3D Artery Anatomy from Angiograms

The reconstruction of 3D artery anatomy includes three steps: artery extraction from fluoroscopy angiograms, imaging geometry calibration, and vessel point correspondences & 3D vessel reconstruction.

**Artery extraction.** A deep learning model[9] was used to extract the coronary arteries on fluoroscopy angiograms. The extracted artery contours were shown in Figure 1 B&F. Based on the extracted artery contours, a morphology thinning based algorithm[10,11] was used to skeletonize the extracted artery trees and an edge-linking algorithm[12] was then applied to link the separate skeleton pixel points, where adjacent skeleton pixel points were linked together to form vessel segments till encountering edge junctions or endpoints. An interactive tool was developed to select the vessel segments to construct complete artery centerlines. Manually drawn segments were involved when the deep learning model performed poorly and therefore failed artery skeletonization. Accordingly, the centerlines on the primary and secondary projection views were extracted and paired (Figure 1 C&G). The topology of the artery anatomy was then automatically established and the bifurcations between the arteries were automatically identified. The radii of vessels were obtained by computing the distance between the centerlines and the outer contour of corresponding arteries.

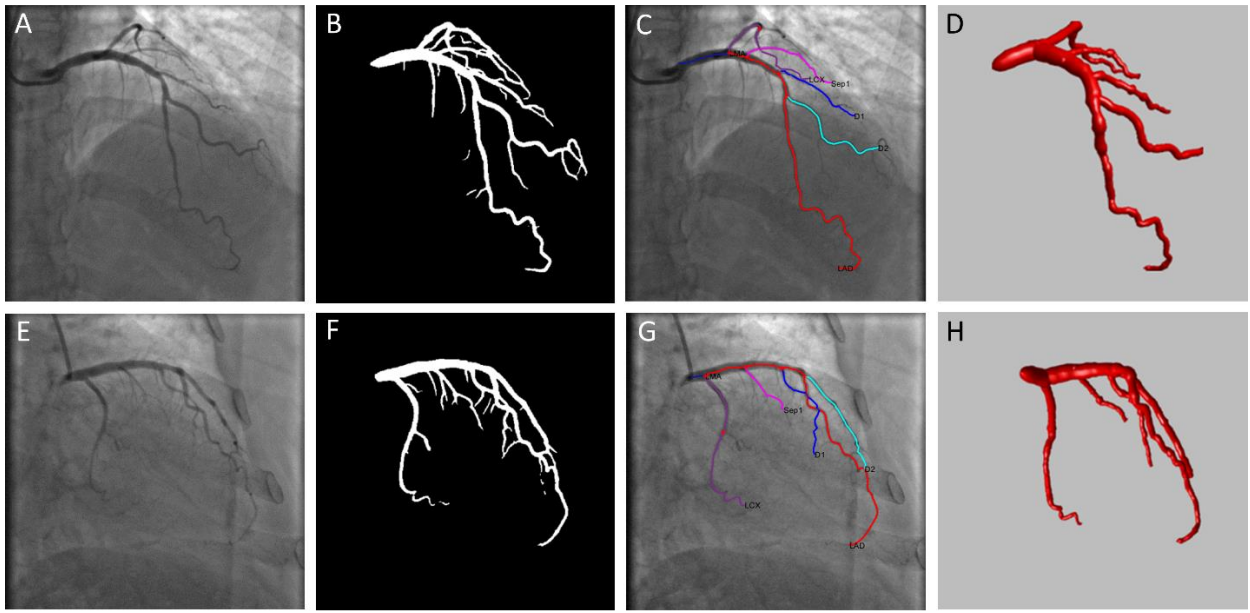

Figure 1. Reconstruction of 3D fluoroscopy artery anatomy. (A) and (E) Selected angiograms from the primary and secondary projection views; (B) and (F) extracted arteries; (C) and (G) artery skeletonization and generation of topology landmarks (red stars) on the angiograms; (D) and (H) reconstructed 3D artery anatomy.

**Imaging geometry calibration.** Imaging geometry parameters defining the projections and orientations for the primary and secondary views are key factors for 3D artery reconstruction. However, these parameters obtained from DICOM header may not be able to accurately define the spatial relationship of these two views because of several uncertainties, such as unknown image skew parameters, table translation between image acquisitions, and device assembly tolerances. A calibration algorithm based on multi-objective optimization was developed to optimize these parameters and explained below (steps i and ii).

i) A mathematical model was first developed. As shown in Figure 2, a spatial bifurcation point $Q_i$ is projected at an intersection point $q_{1,i}$ on the primary projection plane and at an intersection point $q_{2,i}$ on the secondary projection plane. Based on the principles of X-ray angiography and pinhole camera models[13,14], projection matrix mapping spatial point $Q_i$ to projection points ($q_{1,i}$, $q_{2,i}$) was derived. In the coordinate system of primary view, projection matrix $P_1$ can be expressed as in $Eq.\,1$, where $SID$ is the distance between X-ray source and center of detector, s is the skew parameter in radial direction, $u_c$ and $v_c$ are the center coordinates of detector. Since the transformation from the primary to the secondary projection systems can be defined as a rotation $R$ and translation $t$, the projection matrix $P_2$ can be formulated as in $Eq.\,2$. With a preset skew parameter s, all the geometry parameters in the equations can be initialized, though they may be not precise, by the parameters from DICOM header. Therefore, given two projection points $q_{1,i}(u_1, v_1)$ and $q_{2,i}(u_2, v_2)$, the spatial point $Q_i(x_i, y_i, z_i)$ can be obtained by solving an over-determined equation created by the combination of $Eq.\,1$ and $Eq.\,2$, as shown in $Eq.\,3$, where $p_a^{bT}$ is the $bth$ row of the projection matrix $P_1$ or $P_2$, $a$ = [1, 2], $b$ = [1, 2, 3].

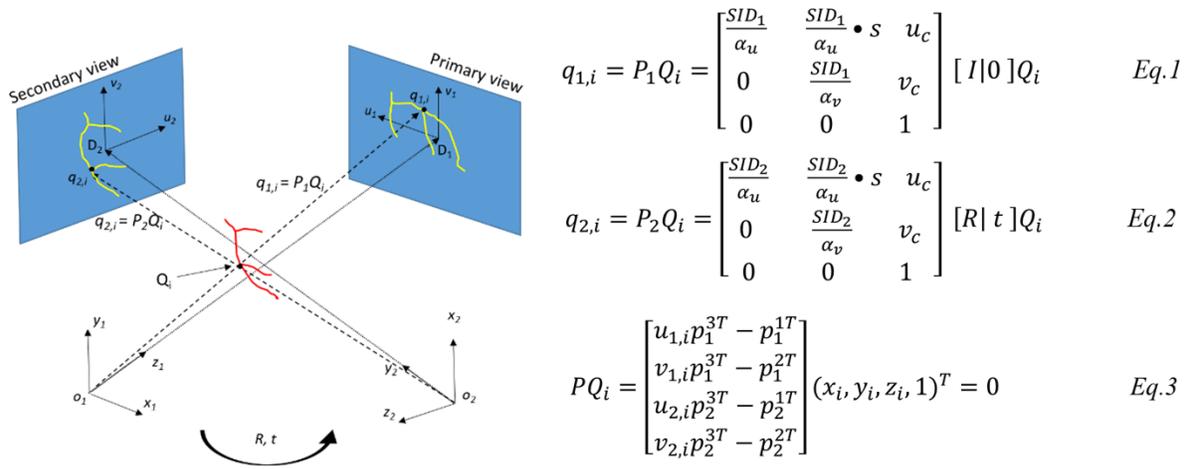

$$q_{1,i} = P_1 Q_i = \begin{bmatrix} \frac{SID_1}{\alpha_u} & \frac{SID_1}{\alpha_u} \bullet s & u_c \\ 0 & \frac{SID_1}{\alpha_v} & v_c \\ 0 & 0 & 1 \end{bmatrix} [I|0] Q_i \qquad Eq.1$$

$$q_{2,i} = P_2 Q_i = \begin{bmatrix} \frac{SID_2}{\alpha_u} & \frac{SID_2}{\alpha_u} \bullet s & u_c \\ 0 & \frac{SID_2}{\alpha_v} & v_c \\ 0 & 0 & 1 \end{bmatrix} [R|t] Q_i \qquad Eq.2$$

$$PQ_i = \begin{bmatrix} u_{1,i} p_1^{3T} - p_1^{1T} \\ v_{1,i} p_1^{3T} - p_1^{2T} \\ u_{2,i} p_2^{3T} - p_2^{1T} \\ v_{2,i} p_2^{3T} - p_2^{2T} \end{bmatrix} (x_i, y_i, z_i, 1)^T = 0 \qquad Eq.3$$

Figure 2. Mathematical model of fluoroscopy angiography system. $q_{1,i}$ and $q_{2,i}$ are the projection points of a 3D arterial bifurcation on the primary and secondary planes. Rotation $R$ and translation $t$ establish the relationship of the primary and secondary coordinate systems. $Eq.\,1, 2,$ and $3$ denote the mathematic model.

ii) An objective function was then proposed and optimized to calibrate the geometry parameters. Based on the mathematical model and initial geometry parameters from DICOM header, an objective function containing 15 geometry parameters were created to minimize the following mismatch errors[14]: (1) Euclidean distance between the artery bifurcations and the corresponding back projections of reconstructed 3D bifurcations on each image; (2) difference between the directional vectors defined by artery bifurcations and the corresponding back projections of reconstructed 3D bifurcations on each image. A nonlinear optimization algorithm, Levenberg-Marquardt (LM)[15], was used to optimize the objective function to obtain the calibrated geometry parameters.

**Vessel points correspondence & 3D reconstruction.** With the calibrated parameters, an epipolar constraint-based method[16,17] was used to pair the vessel centerline points on the primary and secondary images. Given a point on one of the images, there should be at least one corresponding point lying on the epipolar line on the other image. To avoid multiple correspondences, a dynamic programming-based method was used to find the optimal corresponding point. This method minimizes the error defined by the distance of the corresponding point from epipolar line. After establishing the correspondence of the centerline points on the primary and secondary images, 3D artery centerlines were reconstructed using the mathematical model, and then the artery surface was meshed with quadrangles to reconstruct 3D artery anatomy, as shown in Figures 1 D&H.

### 2.1.2. Evaluation of Artery Reconstruction

A computer simulation was implemented to evaluate the accuracy of the artery reconstruction algorithm. Fluoroscopy angiograms were simulated using GATE simulator[18] and X-CAT phantom[19]. In GATE environment, the geometry of Philips system for human body was first set up. Left coronary arteries (LCA) & right coronary arteries (RCA) phantoms generated by X-CAT were then loaded into the system. Two regular views of LCA were simulated, which are LAO45˚ & CRA30˚ (for checking left anterior descending (LAD) artery and its branches) and RAO30˚ & CAU35˚ (for checking left circumflex (LCX) artery and its branches). For RCA, a regular view from LAO1˚ & CRA29˚ for checking distal RCA (posterior descending artery (PDA) and posterolateral branch (PLB)) was simulated, another view from RAO33˚ & CAU5˚ for checking middle RCA was simulated. All the simulated images have a pixel size of 0.34 mm and a resolution of 512×512.

With the simulated angiograms (Figure 3A, B, E, F), LCA and RCA centerlines were reconstructed using the proposed reconstruction algorithm. The 3D LCA and RCA centerlines from X-CAT phantoms were extracted using a 3D thinning algorithm[20]. In order to evaluate the accuracy of the reconstruction algorithm, the mean distances of reconstructed centerlines from the corresponding phantom centerlines (ground truth) were paired and computed, as shown in Figure 3C, D, G, H.

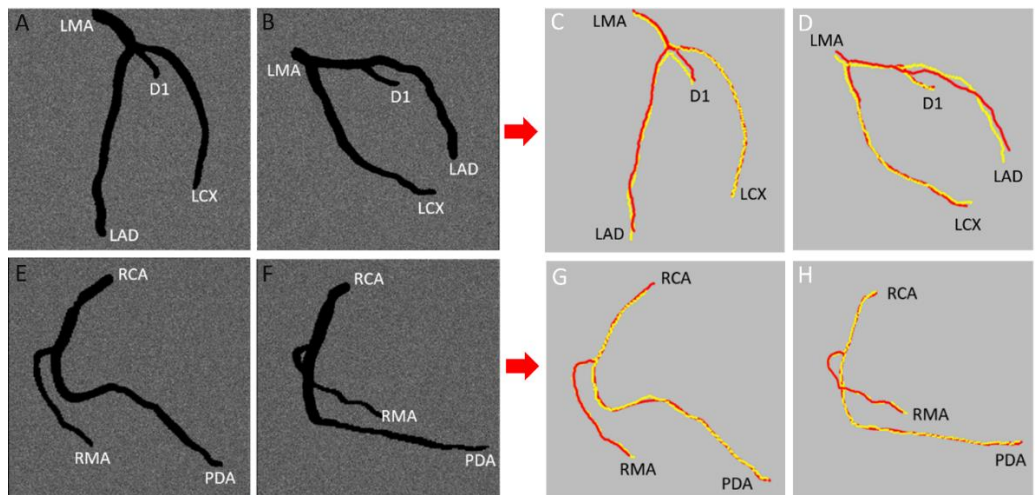

Figure 3. Computer simulation of fluoroscopy angiography and comparison between reconstructed centerlines and the ground truth centerlines extracted from X-CAT phantom. (A) and (B) are simulated LCA fluoroscopy angiograms from LAO45˚ & CRA30˚ and RAO30˚ & CAU35˚; (C) and (D) are the comparisons between reconstructed LCA artery centerlines (red lines) and X-CAT phantom centerlines (yellow lines); (E) and (F) are simulated RCA angiograms from LAO1˚ & CRA29˚ and RAO33˚ & CAU5˚; (G) and (H) are the comparisons between RCA centerlines (red lines) and X-CAT phantom centerlines (yellow lines).

## 2.2 SPECT Image Processing

### 2.2.1. LV Epicardial Surface Extraction from SPECT Images

A graphical user interface was developed to identify LV parameters including LV center, apex, base, anterior and inferior grooves (yellow and green arrows in Figure 4A). Once the parameters were determined, a dynamic programming-based (DP) algorithm[21] was used to extract LV epicardial surface from SPECT images. This algorithm first transformed long-axis SPECT images from Cartesian to polar coordinates and then calculated the gradients of the polar image by the differences in radial direction. LV epicardial contour in the polar image was identified via searching for the maximal gradients using the DP algorithm and thereafter transformed back to Cartesian coordinates. The obtained LV epicardial sampling points were triangulated and then smoothened using a triangulation mesh smoothing algorithm[22]. The surface was rendered with myocardial perfusion data, as shown in Figure 4B. After extracting the LV epicardial surface, the anterior and inferior grooves were generated and used as landmarks for initial alignment of 3D artery anatomy and LV epicardial surface.

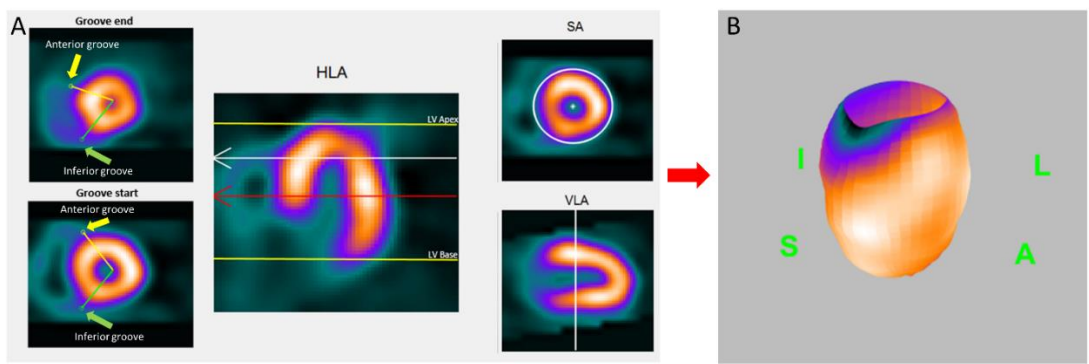

Figure 4. User interface to identify LV parameters and landmarks. (A) Identification of LV parameters. LV center, apex, and base were used to extract LV epicardial surface. Anterior and inferior grooves were used to generate landmarks. (B) Extracted LV epicardial surface from SPECT images.

### 2.2.2. Evaluation of LV Surface Extraction

The accuracy of LV surface extraction was evaluated via a computer simulation. GATE simulator and X-CAT phantom were applied to simulate nuclear images. A Siemens ECAT system for the human body was built in GATE, and a heart phantom generated with X-CAT was loaded into the simulation system. Standard physics processes were included and standard digitizer processing module was set up. Energy resolution was set to 0.26 at 511 keV with Gaussian blurring. The lower and upper bounds of the energy window were initialized to 350 keV and 650 keV, respectively. The source of the simulation was specified by the preset activity values of the organs in the heart phantom. Energy type was set as Mono with 511 keV gamma particles emitted "back-to-back".

With the simulated coincidence data, ordered subsets expectation maximization (OSEM) algorithm packaged by an open-source software(OMEGA)[23] was used to reconstruct nuclear images using 8 subsets and 3 iterations. The reconstructed nuclear images have a voxel size of 3.2 $mm^3$. Butterworth filter was then used to post-process the reconstructed images using a lowpass of 5, a highpass of 50, and an order of 5. The nuclear images after processing were shown in Figure 5A.

The nuclear images were then processed using our DP-based approach to extract the LV epicardial surface (DP-LV surface), as shown in Figure 5C. For comparison, an experienced operator who was blinded from DP-LV surface manually extracted the LV epicardial surface (ME-LV surface) (Figure 5B) using a semi-automatic segmentation tool[24]. The sampling points of DP-LV and ME-LV surfaces were paired (Figure 5D). The mean distance of DP-LV from ME-LV (ground truth) epicardial surfaces was computed to evaluate the accuracy of the epicardial surface extraction algorithm.

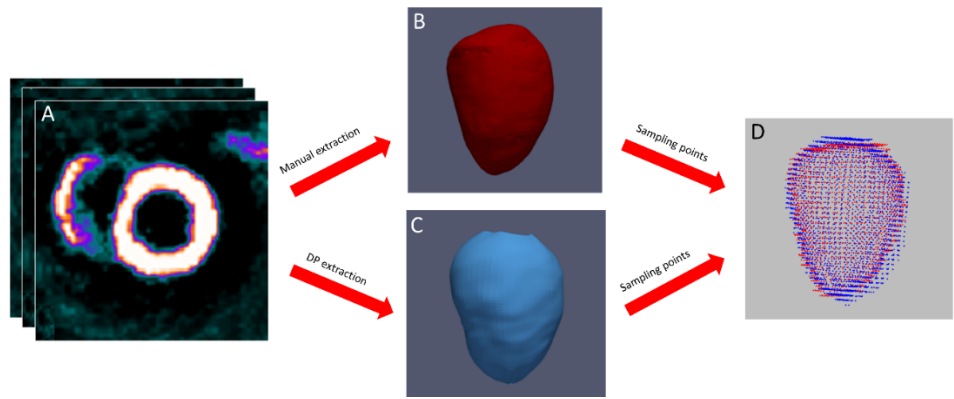

Figure 5. Simulation of nuclear images and LV epicardial surface comparison. (A) Simulated nuclear images using GATE; (B) manually extracted LV epicardial surface; (C) LV epicardial surface extracted by DP-based approach; (D) sampling points comparison between (B) and (C).

**2.3 Image Fusion**

**2.3.1. Fusion between 3D Artery Anatomy and SPECT LV Epicardial Surface**

Three steps were implemented to complete the 3D fusion: 1) landmark-based initial alignment, 2) fine registration using S-ICP, and 3) vessel-surface overlay.

**Landmark-based initial alignment.** According to the characteristics of coronary anatomy[25], LAD travels in the anterior interventricular groove, proximal LCX travels in the left atrioventricular groove, and PDA travels in the inferior interventricular groove. The grooves obtained from SPECT images (section 2.2.1) were used as landmarks (as the white arrow shown in Figure 6A) to complete rough alignment of arteries and LV surface. A cost function was created by minimizing the sum of squared distance between the following three curve pairs: a) between LAD and anterior interventricular groove, b) between proximal LCX and LV base, c) between PDA and inferior interventricular groove, as shown in Figure 6A.

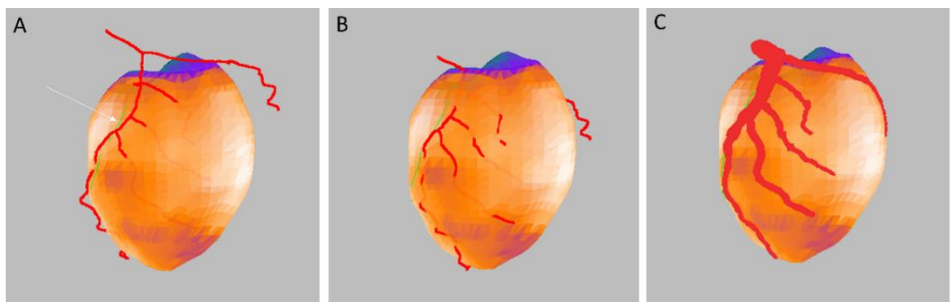

Figure 6. Fusion of artery anatomy and LV epicardial surface. (A) Rough registration by landmarks (green line as illustrated as white arrow); (B) fine registration using S-ICP; (C) vessel overlay and vessel contour rendering.

**S-ICP fine registration.** Since it is quite difficult to match the time points between MPI and angiography during image acquisition, the scale size of reconstructed 3D artery anatomy may be different from the SPECT epicardial surface. Hence, deformable registration is needed to enhance the precision of the 3D fusion. S-ICP, a non-rigid registration algorithm[26,27], was implemented to refine the initial alignment by landmark-based approach. It introduced a scaling factor into standard ICP to form a quadric constraint optimization problem concerning a transformation with respect to scale S, rotation R, and translation t. Two steps were iteratively executed to solve this optimization problem. The first step was to create correspondences between LV epicardial sampling points and artery centerline points in current status. A Delaunay triangulation based algorithm[28] was used to create the correspondence by searching in epicardial sampling points which are closest to the artery centerline points. The second step was to optimize an objective function that minimizes the distance of artery centerline points from the corresponding points in epicardial sampling points. Singular value decomposition (SVD) based method[29] was used to optimize the objective function. Therefore, the transformation parameters (R, S, and t) were obtained until the iteration reaches a preset threshold. Figure 6B shows the result of S-ICP fine registration.

**Vessel-surface overlay.** After the fine registration by S-ICP algorithm, all the arteries were overlaid onto the SPECT LV epicardial surface using a vessel-surface overlay algorithm[30], and then artery contours were created using quadrangles as shown in Figure 6C.

### 2.3.2. Evaluation of the 3D Fusion

The accuracy of the 3D fusion was evaluated using both computer simulation and real patient data.

**Computer simulation.** The artery anatomies from simulated angiograms (section 2.1.2) and DP-LV surface (section 2.2.2) were fused using the 3D fusion approach. The LV epicardial surface extracted from the X-CAT phantom was manually registered with the DP-LV surface, and then the phantom arteries were overlaid onto the DP-LV surface. Though both PDA and PLB in the RCA system travel on the LV surface, only PDA exists in the X-CAT phantom, so PDA was overlaid on the DP-LV surface. The mean distances of fluoroscopy arteries from phantom arteries (ground truth) were computed to technically evaluate the accuracy of the 3D fusion, as shown in Figure 7.

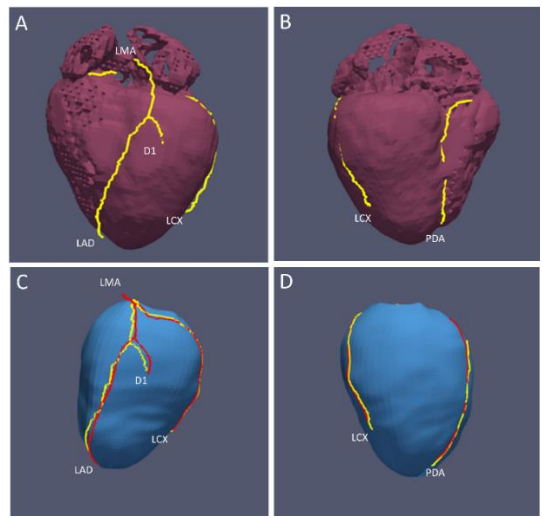

Figure 7. Comparison of fluoroscopy and phantom centerlines. (A) and (B) are artery centerlines and LV&RV epicardial surfaces extracted from X-CAT phantom. (C) and (D) coronary artery centerlines from fluoroscopy angiogram (red lines) and X-CAT phantom centerlines (yellow lines) overlaid on SPECT epicardial surface. LMA = left main artery; LAD = left anterior descending artery; D1 = the first diagonal artery; LCX = left circumflex artery; PDA = posterior descending artery.

**Real Patient Data.** Thirty patients (21 males, and age = 63.0±8.68 years) were retrospectively enrolled from The First Affiliated Hospital of Nanjing Medical University. All patients had either stable or exertional angina before they underwent SPECT MPI, FA, and CT angiography. It is noted that 19 of the 30 patients did not show RCA abnormality so each of them only took one RCA angiogram. This study was approved by the ethics committee of The First Affiliated Hospital of Nanjing Medical University.

Fluoroscopy angiograms and SPECT images in 30 patients were integrated using the 3D fusion approach. Their CT angiograms were manually processed by experienced operators who were blinded from the fluoroscopy angiograms and SPECT images. They manually extracted major arteries and LV&RV epicardial surfaces on the CT angiograms using open-source software (3D slicer)[31], and then registered the CT LV epicardial surface with SPECT epicardial surface via aligning the landmarks (LV base, frontier, and inferior grooves) on both epicardial surfaces. The transformation parameters of registration were also applied to the extracted CT arteries (LAD, LCX, PDA, PLB, and their branches) which travel on the LV epicardial surface, and therefore the CT arteries were closely aligned to the SPECT epicardial surface. The aligned CT arteries were overlaid onto the SPECT epicardial surface and regarded as the ground truth to evaluate the accuracy of the 3D fusion. Figure 8 is an example illustrating the comparison of the fluoroscopy and aligned CT arteries.

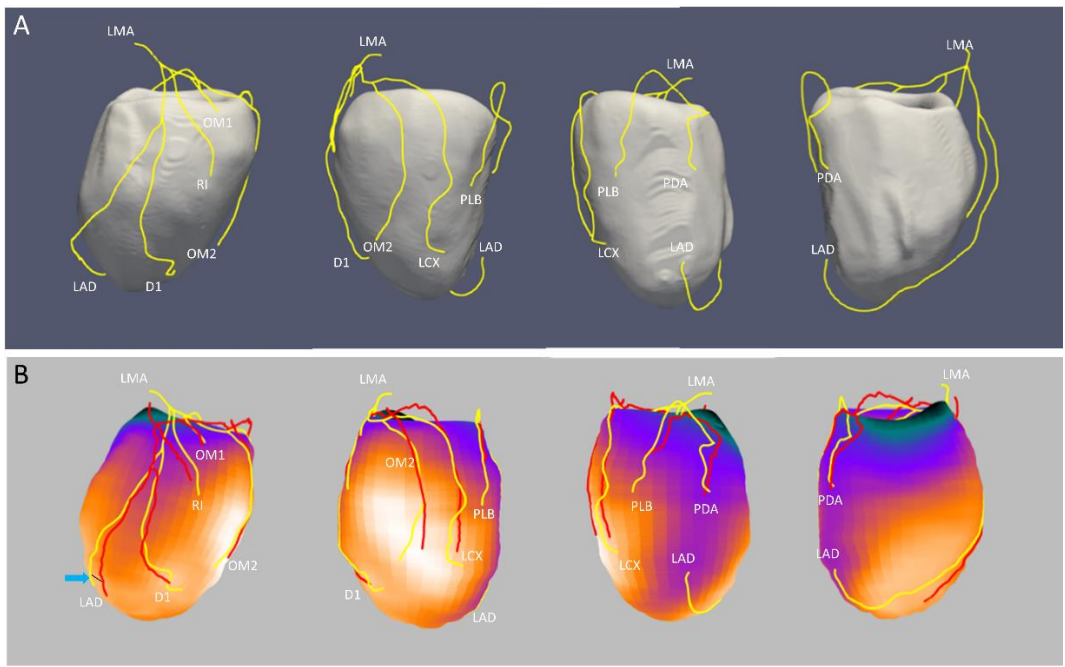

Figure 8. Comparison of fluoroscopy and CT artery anatomy. (A) Coronary arteries (yellow lines) on the CT LV epicardial surface; (B) coronary arteries from CT and fluoroscopy angiograms (red lines) overlaid on the SPECT LV epicardial surface. The mean distance of fluoroscopy and CT arteries on the SPECT epicardial surface, as illustrated by the blue arrow, was used to evaluate the accuracy of the 3D fusion. LMA = left main artery; LAD = left anterior descending artery; RI = ramus intermedius artery; D1 = the first diagonal artery; LCX = left circumflex artery; OM1 = the first obtuse marginal artery; OM2 = the second obtuse marginal artery; PDA = posterior descending artery; PLB = posterolateral branch artery.

**Metrics to Evaluate the Accuracy of 3D Fusion.** Two metrics were used to evaluate the accuracy of 3D fusion. The first one is distance-based mismatch error between CT and fluoroscopy arteries on the SPECT LV epicardial surface. It stands for the mean distance between paired CT and fluoroscopy artery points, as the black line illustrated by the blue arrow in Figure 8. If the fluoroscopy and CT artery lengths were different, the distances were only computed for the paired points. The second metric is a segment-based Kappa agreement rate using the AHA 17-segment model. The segments that both fluoroscopy and artery arteries present were recorded and used to compute the Kappa agreement rate.

## 3. RESULTS

### 3.1. Accuracy of Artery Reconstruction

Table 1 lists the distance-based mismatch errors between simulated fluoroscopy and phantom arteries. In LCA system, left main artery (LMA), LAD, the first diagonal branch artery (D1), and LCX had mismatch errors of 1.97±0.52, 2.32±1.14, 1.99±0.65, and 0.93±0.43 (unit: mm), respectively. A total of 368 fluoroscopy-phantom artery point pairs were evaluated with an overall mismatch error of 1.67±1.07 mm (min: 0.13, max: 4.85). In RCA system, RCA, right marginal artery (RMA), and PDA had mismatch errors of 1.94±1.68, 0.27±0.40, and 0.27±0.43 (unit: mm), respectively. A total of 392 point pairs were evaluated with an overall mismatch error of 1.22±1.06 mm (min: 0, max: 8.12).

| Table 1. Distance-based mismatch errors between simulated fluoroscopy and phantom arteries | | | | |
|---|---|---|---|---|
| **LCA system** | Point pairs | Mean±SD(mm) | Minimum | Maximum |
| LMA | 11 | 1.97±0.52 | 0.65 | 2.50 |
| LAD | 163 | 2.32±1.14 | 0.22 | 4.85 |
| D1 | 34 | 1.99±0.65 | 0.32 | 2.78 |
| LCX | 160 | 0.93±0.43 | 0.13 | 2.08 |
| Overall | 368 | 1.67±1.07 | 0.13 | 4.85 |
| **RCA system** | Point pairs | Mean±SD(mm) | Minimum | Maximum |
| RCA | 157 | 1.94±1.68 | 0.17 | 8.12 |
| RMA | 93 | 0.27±0.40 | 0 | 2.02 |
| PDA | 142 | 0.27±0.43 | 0 | 1.99 |
| Overall | 392 | 1.22±1.06 | 0 | 8.12 |
| Abbreviations: LMA = left main artery; LAD = left anterior descending; D1 = diagonal branch 1; LCX = left circumflex; RCA = right coronary artery; PDA = posterior descending artery; PLB = posterolateral branch. | | | | |

### 3.2. Accuracy of LV Surface Extraction

A total of 4200 DP-ME-LV sampling point pairs were evaluated. The mean distance between DP-LV and ME-LV surfaces (ground truth) was 2.82±1.53 mm (min: 0.02, max: 14.5), which is smaller than the pixel size of the simulated nuclear image (3.2 mm).

### 3.3. Accuracy of 3D Fusion

In the technical evaluation with computer simulation, Table 2 lists the distance-based mismatch errors between simulated fluoroscopy and phantom arteries (ground truth) after registered and overlaid on the DP-LV surface. In the LCA system, LMA, LAD, D1, and LCX had mismatch errors of 3.47±2.17, 2.30±1.74, 2.48±0.42, and 1.18±0.61 (unite: mm), respectively. A total of 368 fluoroscopy-phantom artery point pairs on the simulation LV surface were evaluated for LCA, and overall mismatch error is 1.86±1.43mm (min: 0.04, max: 6.87). In the RCA system, due to the absence of PLB, only PDA was evaluated. PDA had a mismatch error of 2.21±2.25mm (min: 0.05, max: 10.74).

| Table 2. Distance-based mismatch errors between simulated fluoroscopy and phantom arteries on the simulation LV surface | | | | |
|---|---|---|---|---|
| **LCA system** | Point pairs | Mean±SD(mm) | Minimum | Maximum |
| LMA | 11 | 3.47±2.17 | 0.72 | 6.85 |
| LAD | 163 | 2.30±1.74 | 0.04 | 6.87 |
| D1 | 34 | 2.48±0.42 | 1.67 | 3.27 |
| LCX | 160 | 1.18±0.61 | 0.04 | 3.15 |
| Overall | 368 | 1.86±1.43 | 0.04 | 6.87 |
| **RCA system** | Point pairs | Mean±SD(mm) | Minimum | Maximum |
| PDA | 104 | 2.21±2.50 | 0.05 | 10.74 |
| Abbreviations: LMA = left main artery; LAD = left anterior descending; D1 = diagonal branch 1; LCX = left circumflex; PDA = posterior descending artery. | | | | |

In the clinical validation, Table 3 lists the mismatch errors between fluoroscopy and CT arteries on the SPECT surface in 30 patients' data. In the LCA system, the distance-based mismatch error of LMA was 4.91±2.65mm. The mismatch errors of LAD and its branches, ramus intermedius artery (RI), D1, branch of D1 (D1_b1), the second diagonal artery (D2), the third diagonal artery (D3), and the first septal perforator (SEP1), were 3.52±2.80, 3.12±2.54, 3.43±2.95, 3.78±3.40, 3.58±3.13, 5.70±2.13, and 5.17±2.02 (unit: mm), respectively. The mismatch errors of LCX and its branches, the first obtuse marginal artery (OM1), the second obtuse marginal artery (OM2), the third obtuse marginal artery (OM3), and the fourth obtuse marginal artery (OM4), were 4.72±3.40, 3.06±2.42, 3.69±2.76, 4.81±3.12, and 3.88±2.80 (unit: mm), respectively. The overall mismatch error of LCA was 3.84± 3.15mm (min: 0, max: 20.46). In the RCA system, the mismatch errors of PDA, PLB, and the first branch of PLB (PLB_b1) were 5.90±3.92, 4.83±2.89, and 7.80±2.84 (unit: mm). The overall mismatch error of RCA was 5.50±3.64mm (min: 0.11, max: 24.25).

| Table 3. Distance-based mismatch errors between fluoroscopy and CT arteries on SPECT LV epicardial surface | | | | |
|---|---|---|---|---|
| **LCA system** | Point pairs | Mean±SD(mm) | Minimum | Maximum |
| LMA | 545 | 4.91±2.65 | 0.33 | 10.46 |
| LAD | 9005 | 3.52±2.80 | 0 | 19.41 |
| RI | 1104 | 3.12±2.54 | 0.03 | 18.39 |
| D1 | 3608 | 3.43±2.95 | 0 | 20.46 |
| D1_b1 | 132 | 3.78±3.40 | 0.07 | 11.73 |
| D2 | 1696 | 3.58±3.13 | 0.01 | 13.65 |
| D3 | 264 | 5.70±2.13 | 0.12 | 9.05 |
| SEP1 | 130 | 5.17±2.02 | 3.20 | 12.71 |
| LCX | 6192 | 4.72±3.40 | 0 | 18.86 |
| OM1 | 2029 | 3.06±2.42 | 0 | 9.61 |
| OM2 | 1926 | 3.69±2.76 | 0 | 16.97 |
| OM3 | 965 | 4.81±3.12 | 0.02 | 20.25 |
| OM4 | 156 | 3.88±2.80 | 0.09 | 9.81 |
| Overall | 27752 | 3.84±3.15 | 0 | 20.46 |
| **RCA system** | Point pairs | Mean±SD(mm) | Minimum | Maximum |
| PDA | 855 | 5.90±3.92 | 1.02 | 24.25 |
| PLB | 574 | 4.83±2.89 | 0.01 | 21.66 |
| PLB_b1 | 152 | 7.80±2.84 | 2.69 | 13.52 |
| Overall | 1581 | 5.55±3.64 | 0.11 | 24.25 |

Abbreviations: LMA = left main artery; LAD = left anterior descending; RI = ramus intermedius artery; D1 = the first diagonal artery; D1_b1= branch of the first diagonal artery, D2 = the second diagonal artery; D3 = the third diagonal artery; SEP1 = the first septal perforator artery; LCX = left circumflex; OM1 = the first obtuse marginal artery; OM2 = the second obtuse marginal artery; OM3 =the third obtuse marginal artery; OM4 =the fourth obtuse marginal artery; PDA = posterior descending artery; PLB = posterolateral branch artery; PLB_1 = the first branch of posterolateral branch artery.

Table 4 lists the segment-based mismatch error between fluoroscopy and CT arteries on the SPECT surface. In the LCA system, the Kappa agreement rates of LAD and its branches, RI, D1, D1_b1, D2, D3, and SEP1, were 0.87 (95% confidence interval (CI): 0.83-0.92), 0.92 (CI: 0.81-1.03), 0.92 (CI: 0.87-0.98), 1.00 (CI: 1.00-1.00), 0.91 (CI: 0.82-0.99), 1.00 (CI: 1.00-1.00), and 1.00 (CI: 1.00-1.00), respectively. The Kappa agreement rates of LCX and its branches, OM1, OM2, OM3, and OM4, were 0.91 (CI: 0.87-0.96), 0.96 (CI: 0.91-1.01), 0.91 (CI: 0.83-1.00), 0.93 (CI: 0.82-1.02), and 1.00 (CI: 1.00-1.00), respectively. The overall Kappa agreement rate of LCA was 0.91 (CI: 0.89-0.93). In the RCA system, the Kappa agreement rates of PDA, PLB, and PLB_b1 were 0.76 (CI: 0.57-0.95), 0.88 (CI: 0.71-1.05), and 0.73 (CI: 0.37-1.10). The overall Kappa agreement rate of RCA was 0.80(CI: 0.67-0.92).

| Table 4. Segment-based mismatch error between the fluoroscopy and CT arteries on the SPECT LV epicardial surface | | | | | | | |
|---|---|---|---|---|---|---|---|
| **LCA system** | | CT-Y | CT-N | **LCA system** | | CT-Y | CT-N |
| LAD | Fluoro-Y | 115 | 12 | SEP1 | Fluoro-Y | 5 | 0 |
| | Fluoro-N | 12 | 371 | | Fluoro-N | 0 | 29 |
| | Kappa (95%CI) | 0.87 (0.83-0.92) | | | Kappa (95%CI) | 1.00 (1.00-1.00) | |
| RI | Fluoro-Y | 13 | 1 | LCX | Fluoro-Y | 92 | 7 |
| | Fluoro-N | 1 | 138 | | Fluoro-N | 7 | 404 |
| | Kappa (95%CI) | 0.92 (0.81-1.03) | | | Kappa (95%CI) | 0.91 (0.87-0.96) | |
| D1 | Fluoro-Y | 55 | 5 | OM1 | Fluoro-Y | 29 | 1 |
| | Fluoro-N | 5 | 394 | | Fluoro-N | 1 | 275 |
| | Kappa (95%CI) | 0.92 (0.87-0.98) | | | Kappa (95%CI) | 0.96 (0.91-1.01) | |
| D1_b1 | Fluoro-Y | 5 | 0 | OM2 | Fluoro-Y | 24 | 2 |
| | Fluoro-N | 0 | 29 | | Fluoro-N | 2 | 210 |
| | Kappa (95%CI) | 1.00 (1.00- 1.00) | | | Kappa (95%CI) | 0.91 (0.83-1.00) | |
| D2 | Fluoro-Y | 28 | 3 | OM3 | Fluoro-Y | 14 | 1 |
| | Fluoro-N | 3 | 205 | | Fluoro-N | 1 | 120 |
| | Kappa (95%CI) | 0.91 (0.82-0.99) | | | Kappa (95%CI) | 0.93 (0.82-1.02) | |
| D3 | Fluoro-Y | 4 | 0 | OM4 | Fluoro-Y | 2 | 0 |
| | Fluoro-N | 0 | 30 | | Fluoro-N | 0 | 15 |
| | Kappa (95%CI) | 1.00 (1.00–1.00) | | | Kappa (95%CI) | 1.00 (1.00–1.00) | |
| | | | | | | CT-Y | CT-N |
| LCA Overall | | | Fluoro-Y | | | 386 | 30 |
| | | | Fluoro-N | | | 31 | 2222 |
| | | | Kappa (CI) | | | 0.91 (0.89-0.93) | |
| **RCA system** | | CT-Y | CT-N | **RCA system** | | CT-Y | CT-N |
| PDA | Fluoro-Y | 11 | 3 | PLB_b1 | Fluoro-Y | 3 | 1 |
| | Fluoro-N | 3 | 119 | | Fluoro-N | 1 | 63 |
| | Kappa (95%CI) | 0.76 (0.57-0.95) | | | Kappa (95%CI) | 0.73 (0.37-1.10) | |
| PLB | Fluoro-Y | 7 | 2 | | | | |
| | Fluoro-N | 2 | 108 | | | | |
| | Kappa (95%CI) | 0.88 (0.71-1.05) | | | | | |
| | | | | | | CT-Y | CT-N |
| RCA Overall | | | Fluoro-Y | | | 22 | 5 |
| | | | Fluoro-N | | | 5 | 291 |
| | | | Kappa (CI) | | | 0.80 (0.67-0.92) | |
| CI = confidence interval; CT-N = total number of segments without CT arteries; CT-Y = total number of segments with CT arteries; Fluoro-N = total number of segments without fluoroscopy arteries; Fluoro-Y = total number of segments with fluoroscopy arteries; other abbreviations as in Table 3. | | | | | | | |

### 3.4. Processing Time

All the images were processed with a personal computer: Core I5 CPU (2.8GHz), 8 GB memory, and Microsoft Windows 10 operating system. In fluoroscopy angiogram processing, the reconstruction of 3D artery anatomy consumed approximately 13 ± 4s. In the SPECT image processing, the construction of LV epicardial surface required 6 ± 3s. The 3D fusion between them consumed 7 ± 2s.

The interactive identification of artery centerlines on the fluoroscopy angiograms was approximately 4 mins. It is the time barrier since the SPECT LV surface can be extracted before intervention surgery.

## 4. DISCUSSIONS

The primary objective of this study was to develop and validate an approach which integrates 3D fluoroscopy artery anatomy with SPECT LV epicardial surface to guide PCI decision-making. The computer simulation technically evaluated the accuracy of the 3D fusion approach. It showed favorable technical accuracy: artery anatomy reconstruction (mismatch error: 1.67±1.07mm for LCA, 1.22±1.06mm for RCA), epicardial surface extraction (mismatch error: 2.82±1.53mm), and fusion between the artery anatomy and the epicardial surface (mismatch error: 1.86±1.43mm for LCA, 2.21±2.25mm for RCA). Besides, the clinical evaluation in 30 patients showed that 3D fusion had mismatch errors of 3.84±3.15mm for LCA and 5.55 ± 3.64mm for RCA, which is much smaller than the segment size of the AHA 17-segment model (~30 × 30mm$^2$)[32]; the Kappa test showed good agreement rates of the fluoroscopy and CT artery locations on the SPECT epicardial surface: 0.91 for LCA, 0.80 for RCA. Accordingly, the 3D fusion approach showed clinical feasibility to fuse 3D artery anatomy from fluoroscopy angiogram with LV epicardial surface from SPECT for guiding revascularization decision-making.

### 4.1. Clinical Significance of 3D Fusion

SPECT-MPI stress testing is considered a "gatekeeper" prior to invasive angiography and/or PCI in patients with stable CAD. Commonly, revascularization is determined based on visual assessment of a coronary vessel taken in context with perfusion abnormalities described in a written report. Several studies have demonstrated that SPECT guided PCI improves morbidity compared to anatomic assessment or medical therapy alone[33,34].

However, SPECT-MPI guided revascularization without fusion is suboptimal. First, the specificity of SPECT MPI is limited by attenuation artifacts. Second, standard polar map distorts the size, shape, and locations of perfusion defects[35]. Third, vascular territories often overlap and do not necessarily follow standard ascribed distributions. Fourth, although human coronary anatomy is generally similar, each patient's coronary tree is unique with variations of branch vessels and dominance. These limitations lead to 50-60% mismatches between standard segment-based myocardial perfusion territories and the distribution of patient specific anatomic coronary trees[36]. Finally, in patients with multivessel disease, SPECT MPI may not demonstrate perfusion abnormalities in each significant vessel. All these factors decrease the diagnostic sensitivity and specificity and in turn, reduce the utility of SPECT-guided revascularization in clinical practice.

Despite the challenges with SPECT-MPI, our data clearly demonstrates the feasibility of real-time 3D fusion of SPECT-MPI and fluoroscopic coronary angiography. Hypothetically, real-time fused data could influence operator decisions. The techniques described are not limited to SPECT-MPI and can easily be applied to positron emission tomography with coronary flow capacity (PET-CFC) which does not use the typical standardized tomographic, segmentation and polar maps[35]. Point of care fusion of PET-CFC with angiography offers tremendous possible advantages given the mortality benefit and improvement in myocardial blood flow seen with the use of PET-CFC[37,38].

Therefore, the anatomic and physiologic integration, initially with SPECT and subsequently with PET-CFC with FA (the ground truth for evaluation of coronary lesion), offers an opportunity to improve revascularization decisions and outcomes.

### 4.2. Fusion Techniques of Vessel Anatomy and LV Epicardial Surface

Over the past decade, several fusion techniques for coronary vessels and LV surface were developed and validated. These techniques are in three categories: 1) landmark-based method. Zhou et al.[6] and Faber et al.[5] proposed landmark-based methods to integrate LV epicardial surface with 3D coronary vessel anatomy. In both studies, the Landmark-based method can only align the major landmark points, however, the branches and extensions of the major vessels may not be accurately aligned. 2) Standard ICP method. Babic et al.[39] and Toth et al.[8] used standard ICP or Go-ICP to fuse LV epicardial surface with coronary vessel trees. Although these two studies completed the fusion by taking advantage of all the vessel points rather than only landmark points, these fusions are rigid transformation and may fail when two models have scale mismatches caused by the separate image acquisitions at different time points of cardiac beating. Therefore, a non-rigid registration has important advantages. 3) Deep learning-based method. Toth et al.[40] used the imitation learning method to register 2D coronary vessels with 2D projection of CT epicardial surface. Due to the complex overlaps of vessels on 2D coronary angiograms, doctors prefer a 3D artery anatomy fusion with LV surface to better exhibit the stenosis of arteries from any views.

The S-ICP algorithm, in our study, non-rigidly registered 3D coronary artery anatomy with SPECT epicardial surface when scale mismatches existed between them. S-ICP adjusted the scale of 3D artery anatomy up to or down to the optimal scale and then registered it with the SPECT epicardial surface for higher fusion accuracy, which enhances the clinical applicability of 3D fusion. The small distance-based mismatch error and high Kappa agreement rate between fluoroscopy and CT arteries affirmed the accuracy of the 3D fusion approach.

### 4.3. Clinical Applicability

Two essential factors may affect the applicability of the 3D fusion technique. First, for 3D artery reconstruction, the spatial angle gap between the primary and secondary projection views preferably ranges from 45˚ to 145˚. LCA angiography usually meets this condition from standard views by viewing LAD and LCX arteries. RCA angiography also meets this condition in most cases but is limited to the cases of which the spatial angle gaps are out of the range. If this occurs, additional views within the range are needed in order to acquire accurate 3D reconstruction. Second, clear interventricular groove landmarks on the short-axis image (Figure 4) are needed for the initial registration of 3D fusion. Fortunately, these landmarks constantly exist and can be identified for most of the enrolled 30 CAD patients.

Two interactive operations, artery centerline identification on fluoroscopy angiograms and landmark selection on short-axis images, may affect the reproducibility of the 3D fusion approach differently. In the first operation, based on the artery contour from the deep learning model, the extracted centerline segments are usually clear except those with overlaps on angiograms. Manual selection for the clear segments maintains relatively high consistency, which barely impacts the reproducibility. For the centerline segments with complicated overlaps, manually drawn segments for correction vary among operators; however, the reproducibility can still be well guaranteed because 1) overlap is relatively limited compared to the entire artery tree, 2) experienced operators can distinguish the overlaps through observing dynamic cine of coronary arteries from different views, and 3) centerline points from the primary and secondary views that meet epipolar geometry constraints are paired in the 3D artery reconstruction, whereas the incorrectly drawn centerline points by the operators will not be paired. In the operation of landmark selection, a small number of CAD patients who show blurry interventricular grooves on short-axis images, the identification of landmarks among operators may be different. In the 3D fusion approach, the landmarks are used to initialize the S-ICP registration. As mentioned above, S-ICP registers all artery centerlines rather than landmarks with LV surface based on their morphological features. Therefore, the variation in landmarks identification among operators barely impacts final registration.

The clinical validation with the 30 patients confirmed the applicability of the 3D fusion technique. The overall small distance-based mismatch error and high Kappa agreement rate ensure the accuracy of the 3D fusion. The average processing time of 4.5 mins is short compared to the procedural time of PCI (approximately 60 mins), which guarantees the feasibility of this technique.

### 4.4. Limitations

The technical accuracy and clinical feasibility of the 3D fusion approach were tested in a relatively small sample size. Prospective validation in a large population with a control group is needed to establish the clinical usefulness of the technique. Besides, the interactive operations affect the reproducibility of the 3D fusion approach, especially the arteries with complicated overlaps. An improved semantic artery extraction is needed to enhance the reproducibility for broader clinical applications.

## 5. CONCLUSIONS

The developed fusion approach is technically accurate to guide revascularization decision-making and clinically feasible to be used in the catheterization laboratory. There is an opportunity to improve the decision-making and outcomes of PCI in patients with stable CAD.

### ACKNOWLEDGMENTS

This research was supported by a grant from The American Heart Association (Project Number: 17AIREA33700016, PI: Weihua Zhou), a grant from Ochsner Hospital Foundation (PI: Weihua Zhou), and a new faculty grant from Michigan Technological University Institute of Computing and Cybersystems (PI: Weihua Zhou). Besides, Shenzhen Raysight Intelligent Medical Technology, Ltd. provided technical support for the centerline extraction of coronary arteries from CT angiograms.

### CONFLICT OF INTEREST

The authors have no conflict of interest to declare.